# Large thermoelectric response in a diluted ferroelectric system: $Ba_{0.7}Eu_{0.3}Ti_{1-x}Nb_xO_3$

Km Rubi and R. Mahendiran[*]

Physics Department, 2 Science Drive 3, National University of Singapore,

Singapore-117551, Republic of Singapore

## Abstract

We investigated the electrical conductivity, thermal conductivity and thermopower as a function of Nb content ($x$) in $Ba_{0.7}Eu_{0.3}Ti_{1-x}Nb_xO_3$ ($0.001 \leq x \leq 0.10$) in the temperature range $T = 400 – 2$ K. The substitution of Nb destabilizes the ferroelectric insulating ground state of $Ba_{0.7}Eu_{0.3}TiO_3$ and transforms into a paramagnetic metal for $x = 0.1$. Thermopower is negative in the entire composition range ($S = -613$ µV/K at 400 K for $x = 0.001$) and its magnitude decreases with increasing Nb content which suggests doping of electrons into empty Ti-3d($t_{2g}$) conduction band. In this series, the dimensionless figure of merit ($ZT$) increases with temperature for all the compositions and the $x = 0.03$ composition exhibits the maximum $ZT$ ($= 0.12$ at 400 K). The enhanced value of $ZT$ is primarily due to the low thermal conductivity of samples in this series (~ 0.7 to 1 W/(m·K) at 400 K) compared to other potential high temperature n-type thermoelectric oxides such as carrier doped $SrTiO_3$ and $CaMnO_3$. The low thermal conductivity in our compounds most likely arises from heavy $Eu^{2+}$ ion and lattice disorder introduced by $Nb^{5+}$ which scatter phonons effectively.

---

[*] Corresponding author (phyrm@nus.edu.sg)

**Introduction**

Thermoelectric materials which can harvest electrical energy from waste heat as well as cool upon applying electrical voltage could play an important role in growing global energy issues if their efficiency can be improved. The efficiency of thermoelectric devices is determined by a dimensionless figure of merit defined as $ZT = \frac{S^2\sigma}{\kappa}T$, where $S$, $\sigma$, $\kappa$, and $T$ are the Seebeck coefficient (thermopower), electrical conductivity, thermal conductivity and temperature, respectively. To maximize the $ZT$ value of a material, a large $S$, high $\sigma$, and low $\kappa$ are essential. Since these transport characteristics are interrelated, several parameters such as charge carrier concentration, effective mass and thermal conductivities need to be carefully optimized. While some alloys ($Bi_2Te_3$, $Bi_2Sb_3$, PbTe, etc)[1] exhibit essential $ZT$ value ($ZT >1$) required for the practical applications around room temperature, they are expensive, possess toxic element and volatile above ~ 300° C. In this context, transition metal oxides are gaining attention as potential thermoelectric materials for high temperature applications because of their chemically stability, non-toxicity, lower raw materials and manufacturing costs compared to metallic alloys. Among them, large values of thermopower are observed in $NaCo_2O_4$,[2] $Ca_3Co_4O_9$,[3] carrier-doped $CaMnO_3$[4] and $SrTiO_3$[5,6,7,8] despite the fact these samples shows metallic like resistivity with high carrier concentration. La-doped $SrTiO_3$ shows a power factor ($PF = S^2\sigma = 36 \, \mu W/cm \cdot K^2$ at 300 K) comparable to the commercially exploited $Bi_2Te_3$.[5] However, the large thermal conductivity of these compounds ($\kappa$ ~ 9-12 W/(m·K) at $T$ = 300 K) restraints the maximum value of $ZT$ at ~ 0.09.[5]

While numerous papers have appeared in recent years on improving $ZT$ value of carrier doped paraelectric $SrTiO_3$, thermoelectric studies on carrier doped ferroelectric $BaTiO_3$ are very limited. Since the early work of electrical transport and thermopower in single domain $BaTiO_3$ crystal by Berglund and Baer,[9] not much attention was paid to the undoped $BaTiO_3$. Recently, Saijo et al.[10] predicted enhancement of thermopower in $BaTiO_3$ due to ferroelectric distortion that causes anisotropy in effective mass. Available studies indicate that the $\kappa$ value of single crystalline and thin film of $BaTiO_3$ is about 4 W/(m·K) at ~400 K and increases with lowering temperature.[11] While stoichiometric $BaTiO_3$ is highly insulating ($\rho > 10^{12}$ Ω.cm at 400 K), increasing oxygen deficiency causes a decrease in resistivity and even metallic-like behavior.[12] $BaTiO_3$ exhibit a large value of thermopower, but poor electrical conductivity and large thermal conductivity.[13,14] The electrical conductivity of a material is directly proportional to the density and mobility of charge carriers. On the other hand, thermal conductivity relies on the heat conducted by charge carriers ($\kappa_e$) and phonons ($\kappa_l$). While $\kappa_e$ depends on the $\sigma$ via the Wiedemann-Franz law $\kappa_e = \sigma L T$, where $L$ is the Lorentz number, the lattice thermal conductivity ($\kappa_l$) depends primarily on various mechanism of phonon scattering and can be reduced by introducing lattice distortion, defect scattering or nano-structuring. The substitution of much smaller or larger ions for the A or B site of perovskite oxides ($ABO_3$) can increase lattice distortion in the crystal structure that may cause diminution of $\kappa_l$. The effect of the rare-earth ion substitution on the thermoelectric properties of $SrTiO_3$ is already demonstrated and thermal conductivity decreases with increasing atomic mass of the rare earth ion.[7,15,16] Gilbert et al.[17] prepared La-substituted $BaTiO_3$ thin films by metal-organic chemical vapour deposition under different partial oxygen pressure and studied electrical conductivity and thermopower. Muta et al.[18] studied thermoelectric properties of doped $BaTiO_3$-$SrTiO_3$ solid solution and found a maximum ZT of ~0.12 at 400 K in $Ba_{0.3}Sr_{0.6}La_{0.1}TiO_3$. Xiao et al. explored thermoelectric properties of $Ba_{1-x}Eu_xTiO_{3-\delta}$ ($0.1 \leq x \leq 0.9$) series from 300 K to 1123 K and found a maximum ZT of 0.24 at 1123 K in $Ba_{0.1}Eu_{0.9}TiO_{3-\delta}$ sample.[19] The substitution of $Eu^{2+}$ for $Ba^{2+}$ does not dope charge carriers but surprisingly enhances thermal conductivity in the above series. However, electrical conductivity of $BaTiO_3$ can be enhanced by substituting a suitable aliovalent ion for $Ti^{4+}$ sites.

Recently, potential of $Ba_{1-x}Eu_xTiO_3$ series for magnetic cooling in cryogenic temperature was demonstrated.[20] A giant magnetocaloric effect (adiabatic temperature change and isothermal magnetic entropy change upon magnetizing or demagnetizing a sample) occurs in this series as a result of suppression of the spin entropy associated with the seven unpaired 4f electrons of $Eu^{2+}$ ions.[20,21] Upon $Eu^{2+}$ substitution, ferroelectric transition in $BaTiO_3$ also shifts down in temperature and $Ba_{0.7}Eu_{0.3}TiO_3$ becomes ferroelectric below 290 K and exhibits poor electrical conductivity.[22] Partial replacement of $Ti^{4+}$ by $Nb^{5+}$ is expected to dope electrons in the conduction band and hence can improve electrical conductivity of $Ba_{0.7}Eu_{0.3}TiO_3$. In this letter, we report electrical and thermoelectric properties of the electron-doped system, $Ba_{0.7}Eu_{0.3}Ti_{1-x}Nb_xO_3$.

**Experimental details**

$Ba_{0.7}Eu_{0.3}Ti_{1-x}Nb_xO_3$ ($0.00 \leq x \leq 0.10$) polycrystals were synthesized using conventional solid-state reaction method. The stoichiometric amount of $BaCoO_3$, $Eu_2O_3$, $TiO_2$ and $Nb_2O_5$ powders were mixed in the appropriate molar ratio, ground and annealed at 1200° C for 24 hours in a reduced atmosphere (95% Ar and 5% $H_2$) for reducing $Eu^{3+}$ into $Eu^{2+}$. After two consecutive grinding and annealing, the powder was pressed into a pellet and sintered at 1300° C for 24 hours in the same atmosphere. The phase impurity and crystal structure were identified by X-ray diffraction experiment at room temperature. The electrical resistance ($R$), thermopower ($S$) and thermal conductivity ($\kappa$) were measured simultaneously using thermal transport option (TTO) in Physical Property Measurement System (PPMS), Quantum Design, USA. For TTO experiment, the pellet was cut into a rectangular shape with the size of approximate $l \times w \times t$ ~10mm × 4mm × 2mm. The Hall effect experiment for two selected samples $x = 0.05$ and $0.10$ was performed in PPMS using the AC transport option.

**Results and discussion**

The main panel of Fig. 1 shows the powder X-ray diffraction pattern collected at room temperature for $Ba_{0.7}Eu_{0.3}Ti_{1-x}Nb_xO_3$. All samples are in single phase and crystallize in a cubic structure. The lattice parameter ($a$) extracted from Rietveld refinements is displayed in the inset of Fig. 1. The value of $a$ for $Ba_{0.3}Eu_{0.7}TiO_3$ is 3.9746 Å, larger than that for $EuTiO_3$ ($a = 3.9082$ Å)[23] but lower than that for $BaTiO_3$ ($a = 3.996$ Å). The lattice parameter increases rapidly with increasing $x$ from 0.001 to 0.006, but the change is gradual and linear from $x = 0.01$ to 0.10. The increment in $a$ value is due to the larger size of $Nb^{5+}$ cation.

Fig. 2(a) shows the temperature dependence of electrical conductivity ($\sigma$) calculated from the measured resistivity for $Ba_{0.7}Eu_{0.3}Ti_{1-x}Nb_xO_3$ ($0.001 \leq x \leq 0.10$). Resistance of the parent compound $Ba_{0.7}Eu_{0.3}TiO_3$ is more than 1 GΩ at room temperature and we could not succeed to measure it even at 400 K using two-probe method. Our dielectric measurement on $Ba_{0.7}Eu_{0.3}TiO_3$ indicated that the sample becomes ferroelectric at 292 K (not shown here). However, a slight substitution of Nb at Ti site reduces the resistance of sample drastically and we were able to measure the four-probe resistance of $x = 0.001$ down to 2 K. The $\sigma$ value of $x = 0.001$ sample is 0.27 $\Omega^{-1}cm^{-1}$ at 400 K, which decreases gradually with decreasing temperature followed by a rapid decrease below 100 K and reaches $1.5 \times 10^{-6}$ $\Omega^{-1}cm^{-1}$ at 2 K. While samples with $x = 0.001$ to 0.05 show insulating-like behavior ($d\sigma/dT > 0$) in the entire temperature range, $x = 0.1$ turns out to be metallic ($d\sigma/dT < 0$) as shown clearly in inset of Fig. 2(a). The conductivity value at room temperature is enhanced by four orders of magnitude as $x$ increases from 0.001 (0.08 $\Omega^{-1}cm^{-1}$) to 0.1 (320 $\Omega^{-1}cm^{-1}$). The composition driven insulator to metal transition at room temperature occurs because $Nb^{5+}$ substitution for Ti$^{4+}$ dopes 3d$^1$ electrons in Ti:3d(t$_{2g}$) band which otherwise was empty in the undoped parent compound. A doping driven insulator to metal transition was earlier reported in BaTi$_{1-x}$Nb$_x$O$_3$ thin films for $x = 0.2$.[24] The high-resolution energy loss spectroscopy (EELS) study by Shao et al. indicated an increase in the fraction of Ti$^{3+}$ with increasing Nb content in EuTi$_{1-x}$Nb$_x$O$_3$.[25]

We show the temperature dependence of thermopower (S) for all compositions in Fig. 2(b). The sign of S is negative for all samples in whole temperature regime, which indicates that electrons are the majority charge carriers in Ba$_{0.7}$Eu$_{0.3}$Ti$_{1-x}$Nb$_x$O$_3$. For $x = 0.001$, |S| has a large value of 613 µV/K at $T = 400$ K and it decreases as temperature decreases and becomes immeasurable below 140 K due to low electrical conductance of the sample. |S| of $x = 0.003$ decreases gradually and linearly as temperature decreases from 400 K but a rapid steplike decrease occurs between 250 K and 50 K, below which |S| decreases again gradually towards zero value. At $T = 2$ K, the |S| is very small (~2 µV/K). The steplike decrease of |S| is also observed for $x = 0.006$ and up to $x = 0.03$, but the temperature regime for this decrease becomes narrower as $x$ increases. The temperature (T'), where steplike decrease turns on, shifts to the lower value as $x$ goes from $x = 0.003$ (250 K) to 0.03 (200 K). On the other hand, the turn-off temperature (T*) shifts to the higher value (T* = 50, 120, 150 and 160 K for $x = 0.003, 0.006, 0.01$ and 0.03, respectively). The steplike decrease of |S| for $x \leq 0.03$ is possibly due to a discontinuous shift in the Fermi level at a structural transition. A steplike decrease of |S| was also observed in BaTiO$_{3-\delta}$ single and polycrystalline samples due to the orthorhombic-rhombohedral transition at the temperature of 190 K.[13] In contrast to samples with $x \leq 0.03$, |S| for $x = 0.05$ and 0.10 decreases almost linearly with temperature in the entire temperature regime. A linear fit of $S(T)$ is displayed by solid lines in Fig 2(b).

For a metallic or degenerate semiconductor system with the Fermi level in a conduction band, S is predicted to vary linearly with temperature following the equation $S = \frac{-8\pi^2 k_B^2}{3eh^2} m^* T \left(\frac{\pi}{3n}\right)^{\frac{2}{3}} (1 + r)$,[8,26] where r, m*, n, k$_B$ are the scattering parameter, effective mass, charge carrier density and Boltzmann constant, respectively. The n values estimated from the Hall effect experiment are $4.5 \times 10^{20}$ cm$^{-3}$ and $1.8 \times 10^{21}$ cm$^{-3}$ for $x = 0.05$ and 0.10, respectively. The carrier mobility ($\mu$) calculated using the classical Drude's formula ($\sigma = ne\mu$) are 0.58 cm$^2$V$^{-1}$s$^{-1}$ for $x = 0.05$ and 0.77 cm$^2$V$^{-1}$s$^{-1}$ for $x = 0.10$. The n and $\mu$ values in this system are of the same order as those in Sr$_{1-x}$La$_x$TiO$_3$ ($x = 0.05$ and 0.10).[5] The m* value estimated from the linear fit of $S(T)$ data are 1.22 m$_e$ ($x = 0.05$) and 4.09 m$_e$ ($x = 0.10$). The $m^*$ value for $x = 0.05$ is nearly identical to that for same level of La doing in $SrTiO_3$ ($m^* = 1.62\ m_e$)[5].

The temperature dependence of total thermal conductivity ($\kappa$) and electrical contribution of thermal conductivity ($\kappa_e$) are shown in Fig. 2(c) by symbols and solid-lines, respectively. The $\kappa_e$ was estimated using the Wiedemann-Franz law $\kappa_e = \sigma LT$, where $L$ is the Lorentz number = $2.45 \times 10^{-8}$ $V^2K^{-2}$. The $\kappa_e$ is very small for $x = 0.001$ and increases monotonically with increasing $x$. A much larger value of $\kappa$ than $\kappa_e$ confirms the dominance of phonon contribution to the measured thermal conductivity. A maximum in $\kappa(T)$ is observed at the temperature of 190±10 K for all compounds and there is no systematic dependence of the temperature corresponding to this maximum and composition ($x$). Above the maximum, the thermal conductivity is mainly due to phonon-phonon scattering and below the maximum, phonon scattering by defects dominates. A similar behavior of $\kappa(T)$ at low temperatures has been observed for $SrTi_{1-x}Nb_xO_3$[27] and $Sr_{1-x}Y_xTiO_3$[28] systems. What is spectacular about our samples, the value of $\kappa$ at 400K which is in the range of $\kappa \sim$ 0.7 to 1 W/(m·K), much smaller than that of La-doped $SrTiO_3$ single crystals (9 to 12 W/(m·K)).[5]

To understand the electrical conduction mechanism we attempted to fit the $\sigma$ data with formulas for various conduction mechanisms. However, none of the standard mechanisms such as thermal activation model for nearest neighbor hopping (Arrhenius law), variable range hopping or polaronic mechanisms describe the conductivity data for the whole temperature range. The $\sigma(T)$ data for $Ba_{0.7}Eu_{0.3}Ti_{1-x}Nb_xO_3$ ($0.001 \leq x \leq 0.05$) follows small polaron conduction mechanism ($\sigma = \sigma_0 T^{-1}\exp(-E_p/k_BT)$, where $E_p$ is the polaron activation energy) in the high temperature regime. Fig. 3 shows the $\ln(\sigma T)$ versus $1/T$ curves with linear fit for $0.001 \leq x \leq 0.05$. While the curves for $x = 0.001$-$0.03$ deviate from the linear fit below 320 K, the curve for $x = 0.05$ is linear from 400 K-200 K. The $E_p$ as a function of $x$ is shown in the inset of Fig. 3. The $E_p$ decreases with increasing $x$ from 0.001 to 0.01 but increases from $x = 0.01$ to 0.05.

Fig. 4(a), (b) and (c) show the dependence of the $\sigma$, $S$ and $\kappa$ on Nb content ($x$) at $T = 400$ K. While $\sigma(x)$ and $|S|(x)$ show monotonic increase and decrease, respectively, $\kappa(x)$ exhibits a non-monotonic behavior. At $T = 400$ K, the $\sigma$ value increases from 0.27 $\Omega^{-1}cm^{-1}$ to 205 $\Omega^{-1}cm^{-1}$ whereas $|S|$ decreases from 613 μV/K to 63 μV/K with varying $x$ from 0.001 to 0.10. The $\kappa$ value first increases from 0.7 to 0.86 as $x$ increases from 0.001 to 0.006 and then decreases from 0.86 to 0.75 with varying $x$ from 0.006 to 0.03 followed by a rapid increase

with further increase in $x$. For $x = 0.10$, $\kappa$ reaches 0.98 W/(m·K). We believe that the initial increase in $\kappa$ value is due to the increase in $\kappa_e$, but a further decrease is due to the reduction in phonon mean free path with increasing Nb doping.

The variation of the power factor ($PF = S^2\sigma$) with $x$ is shown in Fig. 4(d). The $PF$ increases vastly with increasing $x$ and shows a maximum value (2 µW/cm·K$^2$) for $x = 0.03$. This $PF$ value is much smaller than that of La-doped SrTiO$_3$ single crystals (36 µW/cm·K$^2$)[5] and thin films (39 µW/cm·K$^2$)[29] but comparable to that for La-doped BaTiO$_3$ (4 µW/cm·K$^2$)[18]. Fig. 4(d) displays the composition dependence of the $ZT$ at $T = 300$ K and 400 K. The value of $ZT$ increases with increasing $x$ and exhibits a maximum at $T = 300$ K for $x = 0.01$ but at $T = 400$ K for $x = 0.03$. The maximum values of $ZT$ are 0.04 ($x = 0.01$) at $T = 300$ K and 0.12 ($x = 0.03$) at $T = 400$ K. These $ZT$ values are smaller than $ZT \sim 0.09$ for single crystalline La doped SrTiO$_3$ at $T = 300$K[5] but much larger than the trivalent rare-earth ion doped SrTiO$_3$ polycrystalline samples ($ZT < 0.1$ for Sr$_{0.9}$R$_{0.1}$TiO$_3$, where R = La, Gd, Y etc) at $T = 400$K.[8,16] The $ZT$ value is also much larger than La doped BaTiO$_3$ (~0.05)[30] and EuTiO$_3$ (~0.08)[30] at $T = 400$ K. The increasing trend of $ZT$ with increasing temperature for all samples indicates that these materials are promising for high temperature thermoelectric applications. Since the $PF$ of our compounds is not higher than other rare earth doped SrTiO$_3$, the main reason for the enhancement of $ZT$ in our samples is smaller values of thermal conductivity which most likely arises from the combined effect of the heavier mass of Eu$^{2+}$ (mass = 151.97 g/mol) compared to Ba$^{2+}$ (137.33 g/mol) and lattice disorder created by Nb$^{5+}$ in place of Ti$^{4+}$ ion which reduce phonon mean free path.

**Summary**

In summary, our investigation of electrical and thermoelectric properties of Ba$_{0.7}$Eu$_{0.3}$Ti$_{1-x}$Nb$_x$O$_3$ ($0.001 \leq x \leq 0.10$) in the temperature range of $2$ K $\leq T \leq 400$ K reveals a doping-driven insulator to metal transition at room temperature for $x = 0.1$. The value of $|S|$ decreases monotonically from 613 µV/K at 400 K for $x = 0.001$ to 63 µV/K for $x = 0.1$. At room temperature, $x = 0.01$ sample shows the maximum $ZT$ value of 0.04, however, $x = 0.03$ shows the highest $ZT$ value of 0.12 at $T = 400$ K. The $ZT$ value for $x = 0.03$ is the highest among the carrier doped BaTiO$_3$. Low thermal conductivity of these samples (~0.7 to 1 W/(m·K) at 400 K) is a big advantage and we can expect higher $ZT$ values in single crystal or thin films. Since $ZT$ value increases with increasing temperature, a higher $ZT$ value is expected at high temperatures and hence this series could be interesting for high temperature thermoelectric applications. It will be also interesting to find how $ZT$ varies with fixed Nb$^{5+}$ content but varying $Eu^{2+}$ content. We need to theoretically understand the influence of vibration of $Eu^{2+}$ ions and localized 4f electrons on phonon modes and its consequence on thermal conductivity. In addition, we need to understand what really causes the step-like decrease of thermopower below a certain temperature in low-doped samples.

**Acknowledgements:** R. M. thanks the Ministry of Education, Singapore for supporting this work (Grant no. MOE2014-T2-2-118/ R144-000-349-112).

**Figure Captions:**

**Fig.1** Main panel: Room-temperature X-ray diffraction pattern of $Ba_{0.7}Eu_{0.3}Ti_{1-x}Nb_xO_3$ ($0.001 \leq x \leq 0.10$). Inset: The variation of lattice constant (*a*) as a function of *x*.

**Fig.2** Temperature dependence of (a) electrical conductivity ($\sigma$), (b) thermopower (*S*) and (c) thermal conductivity ($\kappa$). In Fig. (c), symbols represent total thermal conductivity and solid lines represent electrical part of thermal conductivity ($\kappa_e$). Inset: (a) The zoomed view of $\sigma(T)$ for $x = 0.10$.

**Fig.3** Main panel: $\ln(\sigma T)$ versus $1/T$ curves with a linear fit. Inset: Polaron activation energy ($E_\rho$) as a function of *x*.

**Fig. 4** The values of (a) electrical conductivity ($\sigma$), (b) Seebeck coefficient (*S*), (c) thermal conductivity ($\kappa$), (d) thermoelectric power factor (*PF*) and (e) Figure of merit (*ZT*) as a function of Nb contents (*x*) at $T = 400$ K and also at 300 K. Lines are guide to eyes.

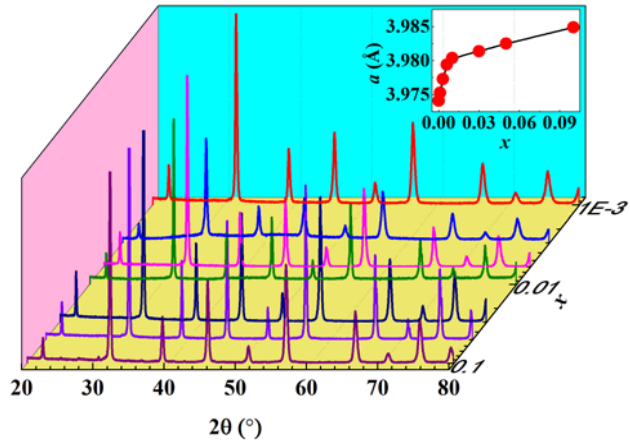

Km Rubi *et al*.

**Fig.1** Main panel: Room-temperature X-ray diffraction pattern of $Ba_{0.7}Eu_{0.3}Ti_{1-x}Nb_xO_3$ ($0.001 \leq x \leq 0.10$). Inset: The variation of lattice constant (*a*) as a function of *x*.

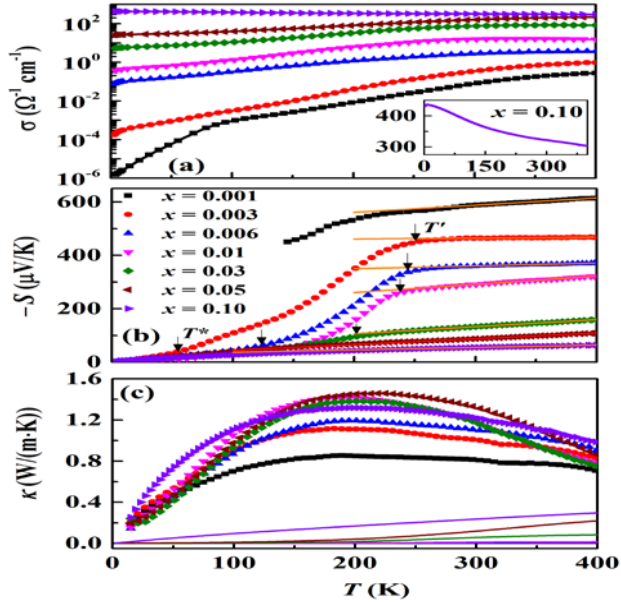

Km Rubi *et al*.

**Fig.2** Temperature dependence of (a) electrical conductivity ($\sigma$), (b) thermopower ($S$) and (c) thermal conductivity ($\kappa$). In Fig. (c), symbols represent total thermal conductivity and solid lines represent electrical part of thermal conductivity ($\kappa_e$). Inset: (a) The zoomed view of $\sigma(T)$ for $x = 0.10$.

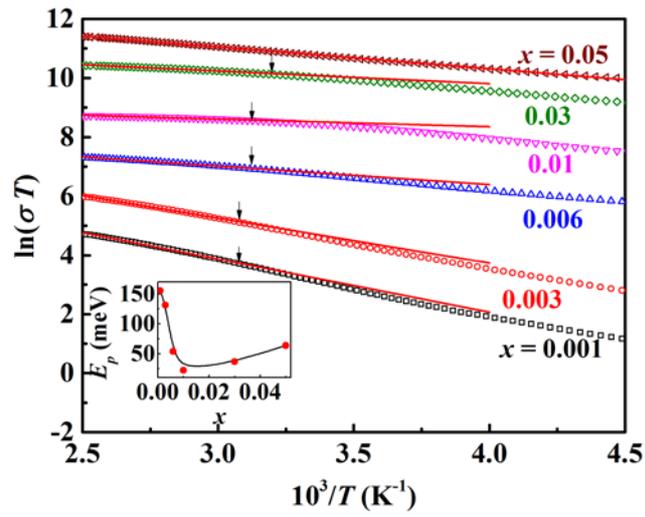

Km Rubi *et al*.

**Fig.3** Main panel: ln($\sigma T$) versus $1/T$ curves with a linear fit. Inset: Polaron activation energy ($E_\rho$) as a function of *x*.

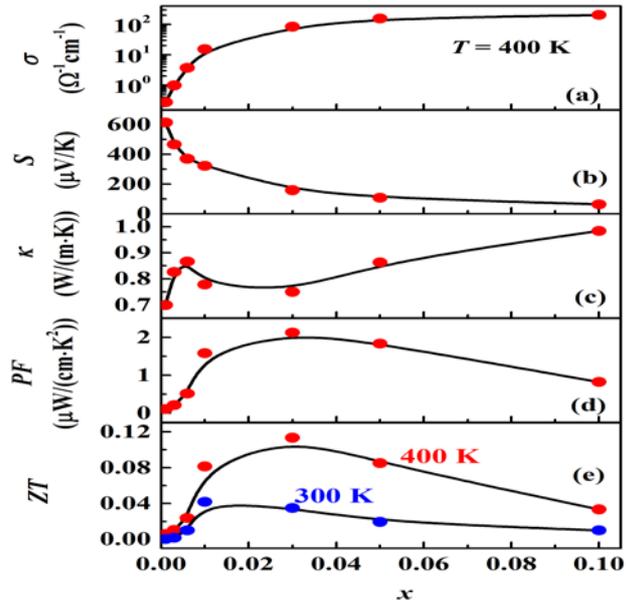

Km Rubi *et al*.

**Fig. 4** The values of (a) electrical conductivity ($\sigma$), (b) Seebeck coefficient (*S*), (c) thermal conductivity ($\kappa$), (d) thermoelectric power factor (*PF*) and (e) Figure of merit (*ZT*) as a function of Nb contents (*x*) at *T* = 400 K and also at 300 K. Lines are guide to eyes.